\documentclass[a4paper,onecolumn,fleqn,usenatbib]{mnras}

\usepackage[T1]{fontenc}
\usepackage{ae,aecompl}
\usepackage{graphicx}	
\usepackage{amsmath}	
\usepackage{amssymb}	

\title[On the enlargement of HZs]{ On the enlargement of habitable zones around binary stars in hostile environments}

\author[Nikolaos Georgakarakos, Siegfried Eggl]{
Nikolaos Georgakarakos,$^{1}$\thanks{E-mail: georgakarakos@hotmail.com}
and Siegfried Eggl,$^{2}$\\ 
$^{1}$New York University Abu Dhabi, Saadiyat Island, P.O. Box 129188, Abu Dhabi, UAE\\
$^{2}$LSST / DiRAC Institute, Department of Astronomy, University of Washington, Seattle, 98015 WA, USA}
\date{Accepted XXX. Received YYY; in original form ZZZ}

\pubyear{2019}

\begin{document}
\label{firstpage}
\pagerange{\pageref{firstpage}--\pageref{lastpage}}
\maketitle

\begin{abstract}
We investigate the hypothesis that the size of the habitable zone around hardened binaries in dense star-forming regions 
increases. Our results indicate that this hypothesis is essentially incorrect. Although certain binary star configurations
permit extended habitable zones, such setups typically require all orbits in a system to be near circular.
In all other cases planets can only remain habitable if they display an extraordinarily high climate inertia. 
\end{abstract}

\begin{keywords}
astrobiology - planets and satellites: atmospheres - planets and satellites: dynamical evolution and stability - planets and satellites: terrestrial planets - (stars:) binaries: general - (stars:) planetary systems
\end{keywords}


In a recent work \cite{2019MNRAS.tmpL..35W} (WP19 hereafter) investigated the orbital evolution of stellar binaries
in densely populated areas such as in star forming regions. By running  N-body simulations they
noticed that a fraction of binaries with initial separations of a few au become even harder with time, i.e. there was a decrease of their semi-major axis. 
This means that the binary components' orbit becomes more likely to feature close approaches, especially when, along  with the hardening of the binary, its orbital eccentricity is excited.  
Following \cite{2013ApJ...777..165K}, WP19 claim 
that the isophote based habitable zone around low mass stars in a binary star system can be enlarged due to the radiative contribution 
of a Solar mass primary, in particular when two distinct isophote based habitable zones overlap.

Although isophote based habitable zones of two stars can theoretically merge in binary star systems, 
the effects of a stronger gravitational environment spell trouble for potentially habitable worlds 
\citep[e.g.][]{eggl2018habitability,pilat2018planetary}.  
The `classical' definition of the habitable zone as given in \cite{1993Icar..101..108K} assumes a planet that moves around a star 
on a circular orbit. That is a plausible assumption for a two body system consisting of the host star and the potentially habitable world. 
In multi-body systems, however, the orbit of a terrestrial planet evolves with time. This is true even if the orbit of the planet is 
initially circular and the whole system is coplanar \citep[e.g.][]{2003MNRAS.345..340G}. 
An immediate consequence of that is that the terrestrial planet will receive varying amounts of insolation 
that may put it temporarily beyond the classical habitable zone. In the worst case, the perturbation on the orbit of the planet
can see it ejected from the system altogether.   
Celestial bodies move according to the laws of gravity, they cannot follow isophotes - lines of equal insolation - in binary star systems \citep{eggl2018habitability}.
Defining habitable zones via isophotes does, therefore, not provide an accurate assessment of the capabilities of a system to host habitable worlds. 

The problem of defining habitable zones for S-type binary star systems including dynamical constraints was addressed in \cite{2012ApJ...752...74E}.
In that work, along with the radiative contribution of the companion star,
the orbital evolution of the terrestrial planet was taken into account. 
In order to account for the effect that orbital mechanics have on the potential habitability of terrestrial planets, \cite{2012ApJ...752...74E}
introduced the concept of Dynamically Informed Habitable Zones (DIHZs).
Depending on the likelihood of the planet to buffer variations in 
insolation without atmospheric collapse, one can distinguish various DIHZs: the Permanently Habitable Zone (PHZ), for instance, is the
area around the host star where the planet stays always within habitable insolation limits, whereas
the Averaged Habitable Zone (AHZ) is defined as the region where the planetary climate can buffer all insolation variations as long as the insolation average remains within habitable limits (a similar idea to that was considered in \cite{2002IJAsB...1...61W}).  Finally,  the Extended Habitable Zone (EHZ) is defined as the area where the planet stays on average plus minus one standard deviation within habitable insolation limits  and it assumes that the planet has limited atmospheric buffering capabilities.
More details about the DIHZs can be found in \cite{2018ApJ...856..155G} and \cite{eggl2018habitability}.
The main conclusion of that work on the subject is that the gravitational perturbations on terrestrial planets 
almost always have an adverse effect on the size of the PHZ. The stronger the gravitational perturbations, the smaller the PHZ. 
Since the PHZ is equivalent to the classical habitable zone for a planet on an evolving eccentric orbit, this means that the harder the binary 
the less likely it is to host habitable worlds in S-type orbits. Of course any change in the orbital eccentricity of the binary also plays a role in that. 
Binary stars on circular orbits are more likely to permit dynamically stable planetary motion with low insolation variance and can, thus, exhibit slightly enlarged PHZs. 

An initially circular orbit of the planet does not necessarily guarantee its habitability, however. In fact, a non-zero initial eccentricity may 
have less of a negative effect on the shrinkage of the classical habitable zone \citep[e.g.][]{2018ApJ...856..155G, eggl2018habitability}, depending on the forced and free eccentricity components.
The main feature that would allow planets to remain habitable inspite of large variations in the incoming starlight is a high climate inertia, i.e. a significant capability to buffer insolation extremes. In such cases, the AHZ can see an extension towards the secondary.

Using an example given in WP19, we demonstrate in the following paragraphs that the system actually becomes less likely to host habitable planes as the binary hardens.
Let a binary consist of a star of mass $m_1=0.63 M_{\odot}$ and effective temperature of 
$T_{eff}=4410K$ and a second Sun-like star of mass  $m_2=0.99 M_{\odot}$ with an effective temperature of 
$T_{eff}=5780K$. The luminosities are calculated from the relation $L/L_{\odot}=(M/M_{\odot})^{3.5}$. Initially the binary has a semi-major axis of $a_{binary}=$6.4 au and an eccentricity of  
$e_{binary}=0.32$. After hardening, the binary has a a semi-major axis of $a_{binary}=$5.4 au and an eccentricity of  $e_{binary}$=0.59. 
By using \cite{1999AJ....117..621H}, WP19 calculate the stability limit around the smaller star at 1.24 au before the interaction and at 0.55 au after the interaction. 
According to them, the 'narrow habitable zone' (runaway greenhouse - maximum greenhouse) for the secondary ranges between 0.32 au and 0.79 au before, and from 0.46 au to 1.26 au after the hardening of the binary.
Hence, a planet on an orbit between 0.46 au and 0.55 au would be stable and inside the enlarged habitable zone according to WP19.

Unfortunately, we were not able to reproduce the results of WP19. First of all,
the stability boundaries that WP19 provide refer to the stable area around the primary star. We suspect that there may have been certain confusion in how to apply the \cite{1999AJ....117..621H} stability criterion. This is of importance since it reverses some of WP19's conclusions
on how hardening affects habitability in binary star systems.  Moreover, WP19 state that the narrow habitable zone before the interaction is 0.32 au - 0.79 au.
If we use \cite{2013ApJ...765..131K} or \cite{2013ApJ...770...82K} (it is not clear which version WP19 have used in their calculation), the habitable zone around the secondary star, rounded to two decimal places, ranges from 0.47 au to 0.84 au or 0.85 au depending on which coefficients are used. The updated version of \cite{2014ApJ...787L..29K} yields a habitable zone of 0.45 au - 0.84 au. That inner limit of 0.32 au is even less that the Recent Venus limit for the single star case which is around 0.35 au. Having a second star in the system, one would expect to see a shift of the inner border of the habitable zone farther away from the host star.


Applying the methodology introduced in \cite{2012ApJ...752...74E} and \cite{eggl2018habitability} to the above scenario (pre and post hardening of the binary) for both 
the 'narrow' (runaway greenhouse and maximum greenhouse limits) and 'empirical' (recent Venus and early Mars limits) habitable zone,  we find that all the DIHZs  in the post-hardening cases either decrease in size or get eliminated entirely due to dynamical instability.  
Figure 1 shows the DIHZs for the pre and post hardening scenarios for a set of orbital eccentricities of the binary star ranging from 0 to 0.8. The orbital eccentricity of the stellar binary clearly affects the size of the DIHZs. This is to be expected as higher binary eccentricity translates into higher planetary eccentricity \citep[e.g.][]{2003MNRAS.345..340G} and hence excursions of the terrestrial planet outside the habitable zone. As expected, the PHZ is the zone that decreases in size the most as the binary eccentricity goes up. As the planetary orbit becomes more eccentric it is more difficult to remain within the classical habitable zone limits at all time.
The area in the $e_{binary}-a_{planet}$ plane where the planetary orbit is stable  also reduces with increasing binary eccentricity. We only see a subtle extension of the habitable parameter space near the outer border of the habitable zone if we assume the binary retains a low eccentricity and the
planet is able to effectively buffer insolation variations.  The specific system discussed in the above example is indicated by a horizontal black line and the corresponding results are presented in Table 1. In the post-hardening case where we use the empirical limits, the part of the PHZ which is not affected by instability is eliminated because it does not satisfy the insolation conditions that would allow an Earth-like planet on a varying orbit to be habitable.  A planet of  $m_p=1 M_{\oplus}$ was considered in the above example. We used \cite{2013ApJ...770...82K} in order to calculate the classical habitable zone limits.

We have verified our results by comparing them to output from the online binary habitable zone calculator {\it BinHab 2.0} of the University of Texas at Arlington which is based on the work of \cite{2014ApJ...780...14C}, \cite{2015ApJ...798..101C} and \cite{2019ApJ...873..113W}. Unfortunately, the calculator can only provide habitable zone and stability limits around the primary. When we calculated the habitable zone borders around the primary star using our model, the results were in good agreement with the numbers obtained by the online tool.  For that calculation of course, we had to force the planetary orbit to remain circular so that our model result could be compared to the one based on the method of Cuntz and collaborators.

In light of the above results we conclude that hardened binary star systems may not be the most favourable type of systems to host habitable planets. 

\begin{figure}
\begin{center}
\includegraphics[width=80mm,height=60mm]{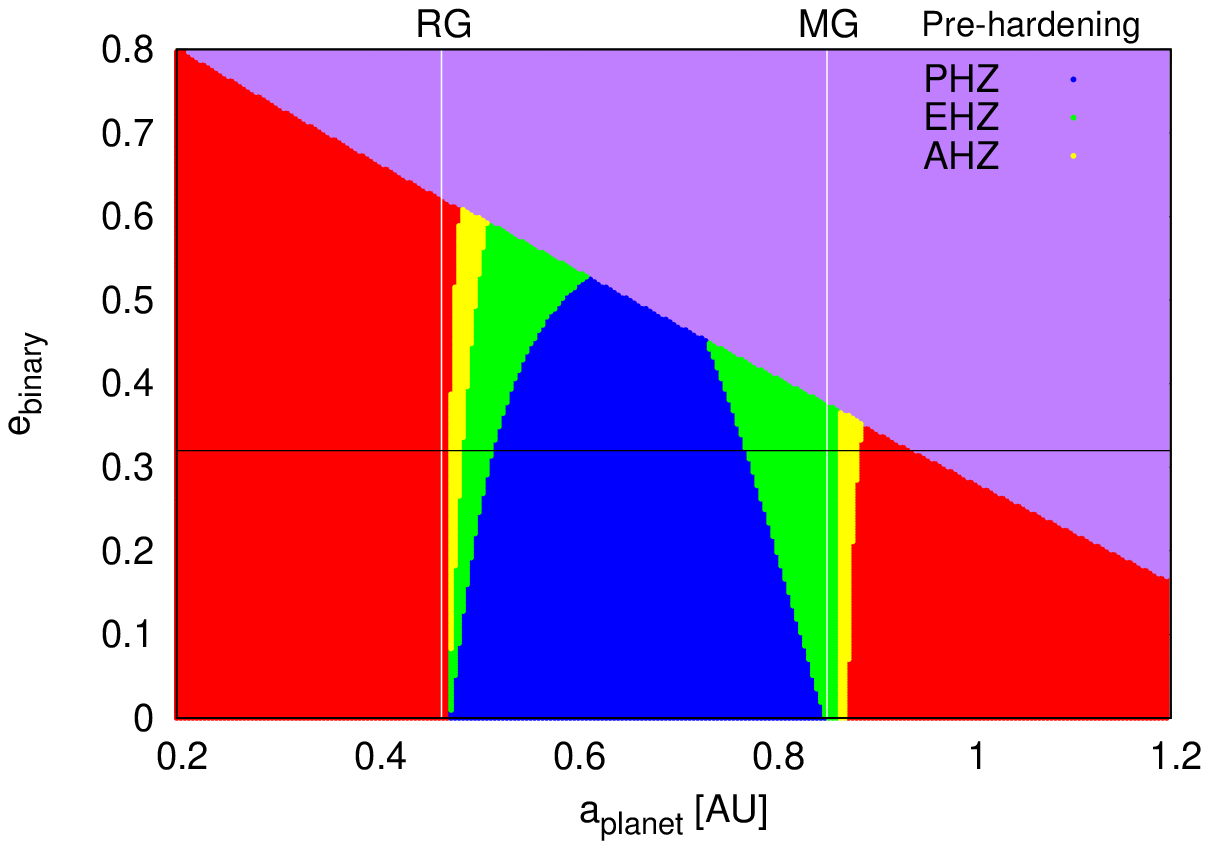}
\includegraphics[width=80mm,height=60mm]{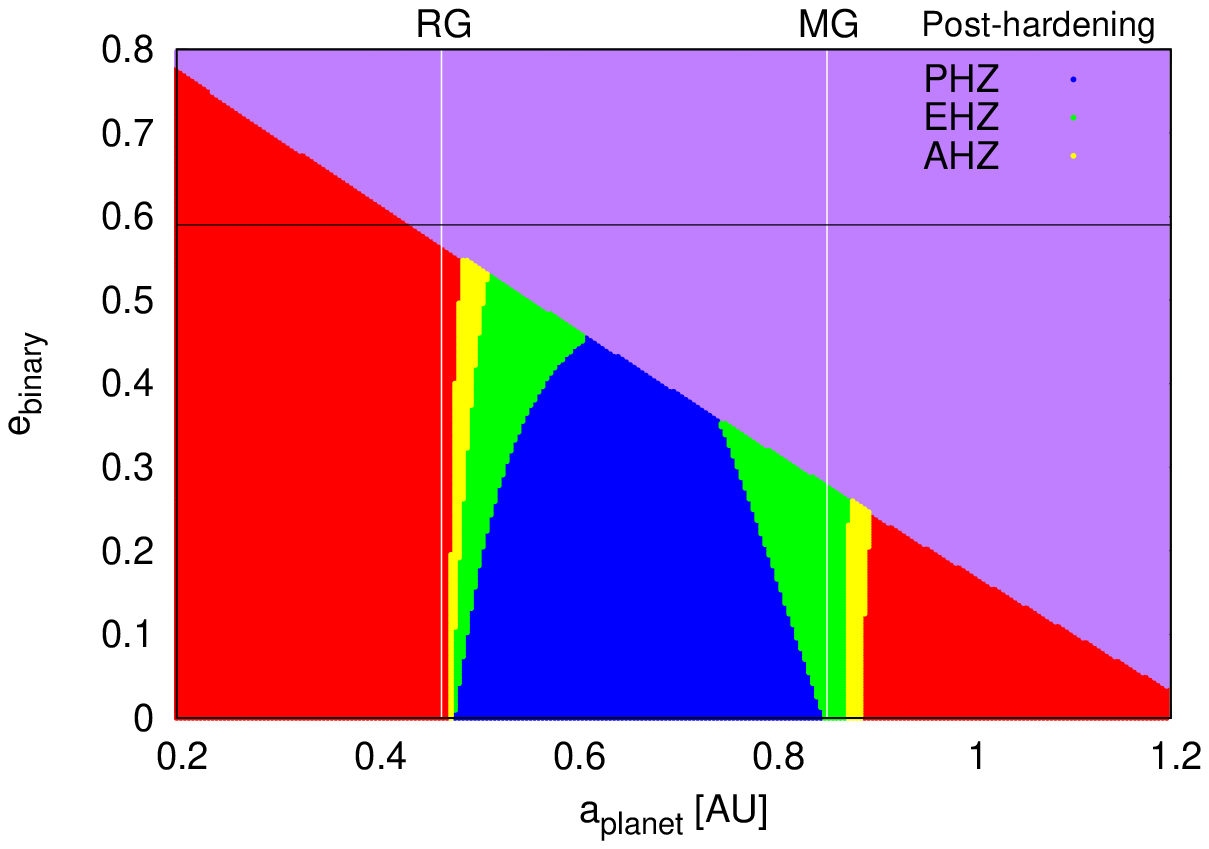}
\includegraphics[width=80mm,height=60mm]{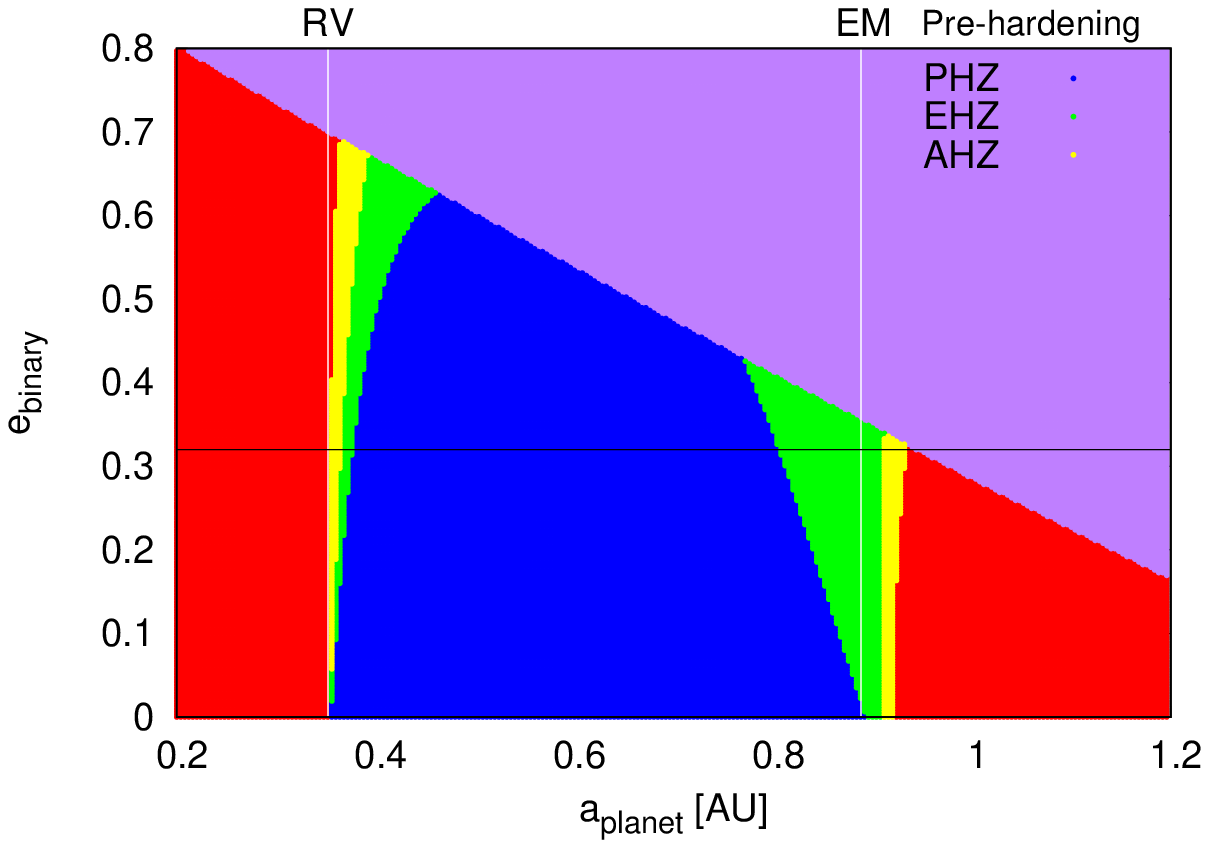}
\includegraphics[width=80mm,height=60mm]{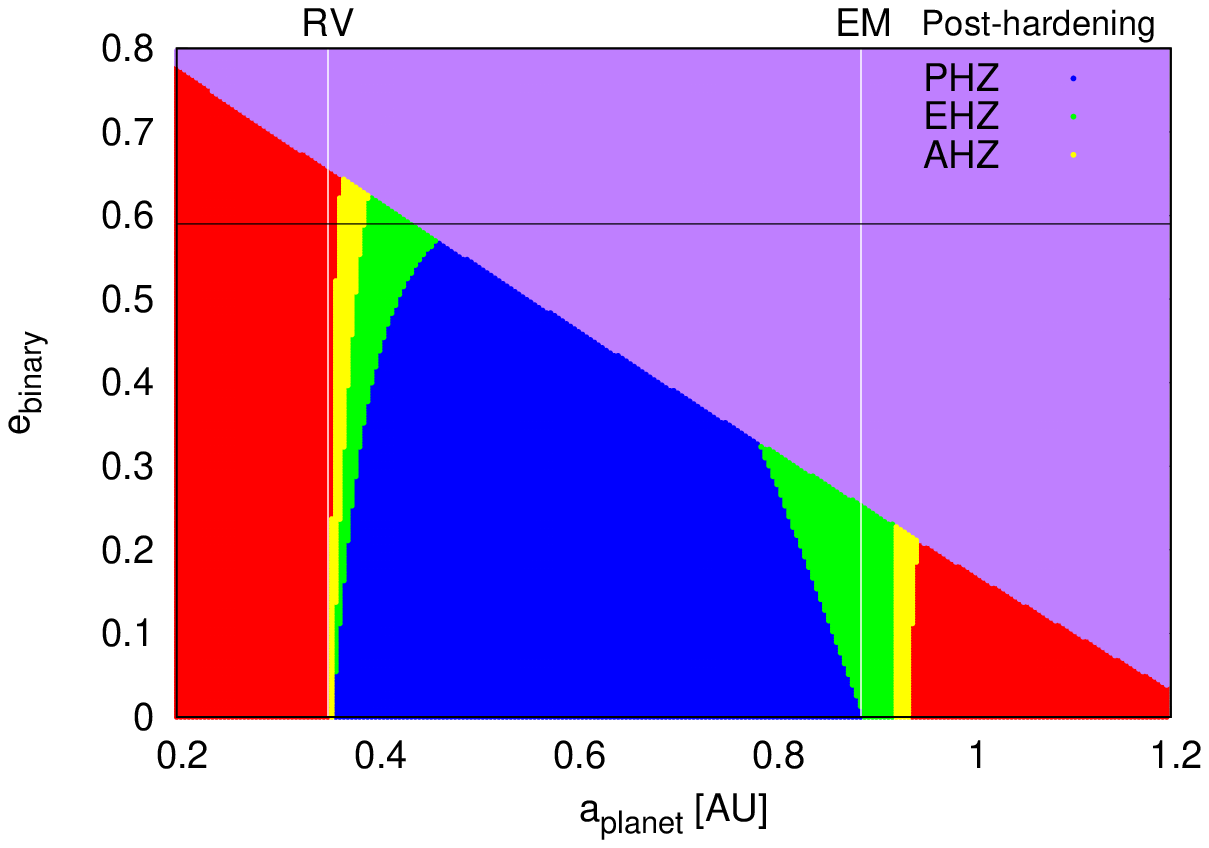}
\caption{ Dynamically Informed Habitable zones for a system with $m_1=0.63 M_{\odot}$, $m_2=0.99 M_{\odot}$ and a planet of  $m_p=1 M_{\oplus}$. The upper row shows the system considering the narrow habitable zone limits  (Runaway Greenhouse RG and Maximum Greenhouse MG), while the bottom row
shows the system considering the empirical habitable zone limits (Recent Venus RV and Early Mars EM). The orbital eccentricity of the binary star discussed in the example in the text is denoted by a horizontal solid black line.  Colours: red - uninhabitable area, magenta - unstable area, blue - PHZ, green - EHZ, yellow - AHZ. 
\label{fig1}}
\end{center}
\end{figure}

\begin{table}
\label{t2}
\caption[]{Habitable zone and stability limits for  a system with $m_1=0.63 M_{\odot}$, $m_2=0.99 M_{\odot}$ and a planet of  $m_p=1 M_{\oplus}$
All the values are rounded to two decimal places.} 
\begin{scriptsize}
\vspace{0.1 cm}
\begin{center}	
{\footnotesize \begin{tabular}{c c c c c c c}\hline
Case &  HZ Limit  & Stability [au]& HZ (single secondary star) [au]& PHZ [au]& EHZ [au]& AHZ [au]\\
\hline
Pre-hardening & Runaway/Max Greenhouse & 0.94 & 0.47 - 0.84 & 0.52 - 0.77 & 0.49 - 0.87 & 0.48 - 0.89\\ 
Post-hardening & Runaway/Max Greenhouse & 0.44 & 0.47 - 0.84 & -  & -  &  - \\ 
Pre-hardening & R. Venus- E. Mars & 0.94 & 0.35 - 0.89 & 0.38 - 0.80 & 0.37 - 0.91 & 0.36 - 0.93\\ 
Post-hardening & R. Venus- E. Mars & 0.44 & 0.35 - 0.89 &  - & 0.39 - 0.43 & 0.36 - 0.43\\ 
\hline
\end{tabular}}
\end{center}
\end{scriptsize}
\end{table}

\section*{Acknowledgements}
S.E. acknowledges support from the DIRAC Institute in the Department of Astronomy at the University of Washington. The DIRAC Institute is supported through generous gifts from the Charles and Lisa Simonyi Fund for Arts and Sciences, and the Washington Research Foundation.  The results reported herein benefited from the affiliation of S.E. with the NASA's Nexus for Exoplanet System Science (NExSS) research coordination network sponsored by NASA's Science Mission Directorate.  
 

\bibliographystyle{mnras}

\bsp	
\label{lastpage}
\end{document}